\documentclass[11pt]{article}
\usepackage{epsfig,sint,macros}
\begin{document}
\begin{titlepage}
\begin{flushright}
  CERN-TH/98-84 \\
  DESY-98-024
\end{flushright}

\vskip 1 cm
\begin{center}
  {\Large\bf  $\Oa$ improvement of lattice QCD with two
      flavors of Wilson quarks }
\end{center}
\vskip 1 cm
\vbox{
\centerline{
\epsfxsize=2.5 true cm
\epsfbox{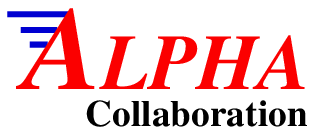}}
}
\vskip 1 cm
\begin{center}
{\large Karl Jansen$^{\scriptscriptstyle a}$
    and Rainer Sommer$^{\scriptscriptstyle b}$}
\vskip 2.5ex
$^{\scriptstyle a}$ 
 CERN, Theory Division\\
CH-1211 Gen\`eve 23, Switzerland\\
\vskip 1.5ex
$^{\scriptstyle b}$
DESY-Zeuthen \\
Platanenallee 6, D-15738 Zeuthen
\vskip 1.5cm
{\bf Abstract}
\vskip 0.7ex
\end{center}
We consider $\Oa$ improvement for two flavor lattice QCD.
The improvement term in the action is computed
non-perturbatively for a large range of the  bare coupling. 
The position of the critical line and higher order
lattice artifacts remaining after improvement are estimated.
We also discuss the behavior of 
the HMC algorithm in our simulations.

\vfill

\begin{flushleft}
  CERN-TH/98-84\\
  DESY-98-024 \\
 March 1998
\end{flushleft} 

\eject

\vfill

\eject

\end{titlepage}

\section{Introduction}

The lattice provides a regularization for QCD, which allows us to study
also non-perturbative aspects of the theory from first
principles. 
In order to implement a lattice regularization, 
a non-vanishing lattice spacing $a$ has to be introduced.  
To achieve the goal of making contact with the physical world, 
it is then unavoidable that results obtained, e.g. by 
numerical simulations, have to be extrapolated
to zero lattice spacing in order to reach 
the continuum limit. 
The rate with which the continuum limit can be approached will then depend
on the amount of contamination of the results by lattice spacing
effects. 

A systematic  approach to reduce discretization errors
is Symanzik's improvement programme for on-shell
quantities 
\cite{Symanzik:1982-Luscher:1985xn}. 
Applying this programme for Wilson fermions,
it turns out that for cancelling the
$\Oa$ effects it is sufficient to add only one new term
into the action as suggested by Sheikholeslami and Wohlert
\cite{impr:SW}. 
The coefficient $\csw$ multiplying this term then has 
to be tuned in such a way that no $\Oa$ effects remain in
on-shell quantities. In order to achieve this, it is necessary 
to determine $\csw$ non-perturbatively by imposing suitable
so-called improvement conditions 
\cite{impr:lett-impr:pap3}.  
For a complete cancellation of all $\Oa$ effects, also the improvement
of various composite fields have to be performed, which introduces 
a not too large number of additional improvement coefficients. 

This programme was successfully applied in the quenched approximation
and the effects of improvement have been verified
\cite{impr:pap3-Edwards:1997nh}. 
Since 
systematic uncertainties
caused by lattice artifacts are strongly suppressed
in the complete $\Oa$ improved theory, 
it is expected
that simulations can be done at larger lattice spacing, reducing in this
way their cost substantially. 
This point of view should hold in particular if we switch from the quenched
approximation to the full theory, including also the effects of
$N_f$ flavors of 
dynamical fermions.  Since simulations with $N_f \ne 0$ are
much more expensive than ones with $N_f =0$, we expect especially in this
situation a considerable gain from performing a non-perturbative 
$\Oa$ improvement. In this paper we therefore will initiate the non-perturbative
computation of the improvement coefficients starting by determining
$\csw$ for $N_f=2$ dynamical flavors of Wilson fermions.
A short account of our work has already appeared in 
\cite{lat97:karl}. 

We emphasize again that, although we expect to be able to accelerate the approach to the
continuum limit by determining the improved theory, nevertheless the
extrapolation to zero lattice spacing has to be performed. Indeed, our results
discussed below indicate that
higher order corrections can still be non-negligible.

\section{Determination of $\csw$}
Our determination of $\csw$ closely follows  \cite{impr:pap1,impr:pap3}.
In these references 
on-shell $\Oa$ improvement and the use of the
\SF in this context are also thoroughly explained. 
In order to make the present paper reasonably 
self-contained, we will introduce the improvement condition explicitly
and  outline the
computation of $\csw$. 
Unexplained notation is taken over 
from \cite{impr:pap1,impr:pap3}. 

\subsection{$\Oa$ improved QCD}

We start from Wilson's formulation of lattice QCD \cite{Wilson}.
The action is the sum of the usual plaquette terms and the
quark action
\begin{equation}
 \Sf=a^4\sum_{x}\psibar(x)(D+m_0)\psi(x),
 \label{e_quark}
\end{equation}
where $a$ denotes the lattice spacing.  The Wilson-Dirac operator,
\begin{equation}
  D=\frac{1}{2}\left\{(\nabstar\mu
  +\nab\mu)\dirac\mu-a\nabstar\mu\nab\mu\right\},
\label{e_Wilson-Dirac}
\end{equation}
contains the lattice covariant forward and backward derivatives,
$\nab\mu$ and $\nabstar\mu$. 
Energy levels and on-shell matrix elements
computed with this action approach their
continuum limits with a rate that is asymptotically linear in the lattice 
spacing. These leading linear terms may be cancelled by adding
one improvement term
to the Wilson-Dirac operator~(\ref{e_Wilson-Dirac}):
\begin{equation}
 D_{\rm impr}=D+\csw\,{{ia}\over{4}}\sigma_{\mu\nu}\widehat{F}_{\mu\nu},
 \label{e_dimpr}
\end{equation} 
where $\widehat{F}_{\mu\nu}$ is 
the standard discretization 
of the field strength tensor~\cite{impr:pap1}.
The coefficient $\csw(g_0)$ is a function of the bare gauge coupling $g_0$
and, when it is properly chosen, 
it yields the on-shell O($a$) improved lattice action, which 
was first proposed by Sheikholeslami and 
Wohlert \cite{impr:SW}\footnote{
For completeness we note that special care has to be taken,
when  massless renormalization schemes
are used. 
This issue is discussed in ref.~\cite{impr:pap1}, but is
not of immediate relevance here, where we want
to perform a non-perturbative
computation of $\csw(g_0)$ for $\nf=2$. }.

When considering matrix elements of local operators, their improvement
has to be discussed as well. Here, we only need the improved isovector
axial current,
\begin{eqnarray}
 (\aimpr)_\mu^a\!\!\!&=&\!\!\!
 A_\mu^a+a\ca\frac{1}{2}(\drvstar{\mu}+\drv{\mu})
       P^a, 
       \label{e_impr_current}
\end{eqnarray}
where
\begin{eqnarray}
  A_\mu^a(x)&=&\psibar(x)\dirac\mu\dirac5{{\tau^a}\over{2}}\psi(x),
  \label{e_current}\\
  P^a(x)&=&\psibar(x)\dirac5{{\tau^a}\over{2}}\psi(x),
\end{eqnarray}
$\drv{\mu}$ and $\drvstar{\mu}$ denote the standard forward and backward lattice
derivative and the Pauli-matrices $\tau^a$ act on the 
flavor indices of the quark fields.
The pseudo-scalar density $P$ needs no $\Oa$ improvement term, 
whereas for the axial current one introduces  
$\ca$ as another improvement coefficient. Current and density defined above 
are not renormalized,
but their multiplicative renormalization is of no 
importance in the following. 
The coefficients $\csw$ and $\ca$ are functions of the bare coupling 
but do not depend 
on the quark mass. Their perturbative expansion
is known to 1-loop accuracy \cite{impr:Wohlert,impr:pap2}, in particular
\bes
 \csw &=& 1 + \csw^{(1)} g_0^2 + \rmO(g_0^4), \quad \csw^{(1)}=0.2659(1) .
 \label{e_csw1}
 \ees
Non-perturbatively, $\csw$ and $\ca$ 
can be computed by imposing suitable improvement conditions.
\subsection{The improvement condition}
The general idea for formulating an improvement condition to fix the
$\Oa$ counter-term in the action is that the $\Oa$ terms violate chiral symmetry.
Hence chiral Ward identities are violated at non-vanishing values of the lattice
spacing. In particular, the unrenormalized PCAC relation 
 \begin{equation}
    \frac12(\drv\mu+\drvstar\mu)
    \langle(\aimpr)_\mu^a(x)\,{\cal O}\rangle=
     2 m \langle P^a(x)\,{\cal O}\rangle 
 \label{e_PCAC_impr}
\end{equation}
contains an error term of order $a$ in Wilson's original formulation, 
which is reduced to $\Oasq$ in the improved theory.
\Eq{e_PCAC_impr} can be taken to define a bare
current quark mass $m$.
Depending on the details of the correlation functions,
such as the choice of the kinematical variables
$\cal O$, the position $x$ and boundary conditions,
one obtains different values of $m$. 
These differences are of order $a$ in general and are
reduced to $\Oasq$ by improvement.
Requiring $m$ to be exactly the same for three choices of the
kinematical variables allows us
to compute the improvement coefficients $\csw$ and $\ca$.

In more detail, we now consider the \SF 
\cite{SF:LNWW-SF:stefan2} with boundary 
conditions
\bes
  \left.U(x,k)\right|_{x_0=0}&=&\exp(aC_k), 
 \quad C_k=\frac{i}{6L}{\rm diag}\left(-\pi,0,\pi\right) 
 \\
  \left.U(x,k)\right|_{x_0=T}&=&\exp(aC'_k), 
 \quad C'_k=\frac{i}{6L}{\rm diag}\left(-5\pi,2\pi,3\pi\right)
\ees
for the gauge fields and boundary conditions for the quark fields as
detailed in ref.~\cite{impr:pap3} (taking  $\theta=0$).
For $\cal O$ we choose 
\begin{equation}
 {\cal O}^a
 =a^6\sum_{\bf y,z}\zetabar({\bf y})\dirac 5{{\tau^a}\over{2}}\zeta({\bf z}),
 \label{e_O}
\end{equation}
where $\zeta$ ($\zetabar$) are  
the ``boundary (anti) quark fields'' \cite{impr:pap1}
at time $x_0 = 0$.
Similarly we use 
\begin{equation}
 {\cal O'}^a
 =a^6\sum_{\bf y,z}\zetabar'({\bf y})\dirac 5{{\tau^a}\over{2}}\zeta'({\bf z}),
 \label{e_Op}
\end{equation}
with the ``boundary fields'' at $x_0=T$.
\Eq{e_PCAC_impr} then leads us to 
consider the correlation functions
\bes
\fa(x_0)&=&-\frac{1}{3}
   \langle A_0^a(x)\,{\cal O}^a\rangle,  \quad 
\fp(x_0)\,\,=\,\,-\frac{1}{3}
   \langle P^a(x)\,{\cal O}^a\rangle \\
\fa'(T-x_0)&=&+\frac{1}{3}
   \langle A_0^a(x)\,{\cal O'}^a\rangle,  \quad 
\fp'(T-x_0)\,\,=\,\,-\frac{1}{3}
   \langle P^a(x)\,{\cal O'}^a\rangle,
\ees
and a  current quark mass is given by
\bes
 \m(x_0)=\r(x_0)+\ca\s(x_0),
\ees
where
\begin{eqnarray}
  \r(x_0)&=&\frac{1}{4}(\drvstar{0}+\drv{0})\fa(x_0)/\fp(x_0),
  \\
  \s(x_0)&=&\frac{1}{2}a\drvstar{0}\drv{0}\fp(x_0)/\fp(x_0) .\label{e_sx0}
\end{eqnarray}
Another mass  $\m'$ is similarly defined in terms of the primed correlation 
functions. Improvement conditions may  be obtained, e.g. by requiring
 $m=m'$ for  
some choice of $x_0$. In order to obtain an improvement condition that  
determines $\csw$, it is, however, advantageous to first eliminate $\ca$, which 
is unknown at this point.
To this end one observes that the combination
\begin{equation}
  \M(x_0,y_0)=
  \m(x_0)-\s(x_0){\m(y_0)-\mprime(y_0)\over\s(y_0)-\sprime(y_0)} \label{e_M}
\end{equation}
is independent of $\ca$, namely
\begin{equation}
  \M(x_0,y_0)=
  \r(x_0)-\s(x_0){\r(y_0)-\rprime(y_0)\over\s(y_0)-\sprime(y_0)}.
\end{equation}
Furthermore, from eq.~(\ref{e_M}) one infers that $\M$ coincides with 
$m$ up to a 
small correction of order $a^2$ (in the improved theory);
$\M$ may hence be taken as an 
alternative definition of an unrenormalized current 
quark mass, the advantage being that we do not need to know $\ca$
to be able to calculate it.

Now we define $\Mprime$ in the same way as $\M$,
with the obvious replacements.
It follows that (amongst others) the difference
\begin{equation}
  \dM=\M\big(\frac{3}{4}T,\frac{1}{4}T\big)-
      \Mprime\big(\frac{3}{4}T,\frac{1}{4}T\big)
\end{equation}
must vanish, up to corrections of order $a^2$, 
if $\csw$ has the proper value.
This coefficient may hence be fixed by demanding 
\bes
\dM=\dM^{(0)}. \label{e_impr_cond}
\ees
Here $\dM^{(0)}$, the value of $\dM$ at tree-level of perturbation theory
in the $\Oa$ improved theory, is chosen instead of zero, in order to cancel a
small tree-level $\Oa$ effect in $\csw$. In this way one ensures that
the values for $\csw$ determined non-perturbatively approach exactly one
when $g_0\to 0$.
To complete the specification of the improvement condition,
we choose $L/a=8,T=2L$ and evaluate \eq{e_impr_cond} for 
quark mass zero, i.e. $\M=0$. 
(We use
$\M$ without arguments to abbreviate 
$\M\big(\frac{1}{2}T,\frac{1}{4}T\big)$ from now on, while $\dM$
as defined in eq.~(2.20) refers to $x_0=\frac{3}{4}T$.) 

The small tree-level lattice artifact in \eq{e_impr_cond} is then evaluated
to 
\bes
a \left. \dM^{(0)}\right|_{M=0,\csw=1} = 0.000277 \,\,{\rm at}\,\, L/a=8.
\ees

Note that in the \SF it is possible to set the quark mass to zero, since
there is a gap in the spectrum of the Dirac operator of order $1/T$.

\subsection{Numerical results for $\csw$}

For a range of bare couplings $g_0$ we want to solve \eq{e_impr_cond}
for $\csw$. The general numerical procedure (which was used in \cite{impr:pap3})
to achieve this is summarized as follows. 
\begin{itemize}
\item[i)]{For fixed parameters $g_0$, $\csw$ and a few
suitably chosen values of the bare quark mass, compute 
$\M$ and $\dM$ and interpolate linearly in $\M$ to find $\dM$ at $\M=0$. }\\[-3.0ex]
\item[ii)]{At fixed $g_0$, repeat i) for a few values of $\csw$ and
find the value of $\csw$ that solves \eq{e_impr_cond} by a 
linear fit in $\csw$.}\\[-3.0ex]
\item[iii)]{Repeat i) and ii) for sufficiently
many values of $g_0$ to be able to find a good
approximant $\csw(g_0)$ for the range of $g_0$ that is of interest.}\\[-3.0ex]
\end{itemize}
Since we are now interested in the theory with two flavors of dynamical fermions,
the calculation of $\M$ and $\dM$ 
for each choice of $m_0,g_0,\csw$
requires a separate Monte Carlo calculation. We did
these calculations with the Hybrid Monte Carlo algorithm. Details of the
simulations and
our error analysis are discussed in \sect{s_HMC}.
Here we note that these simulations are CPU-time intensive and it is
therefore desirable to limit the number of simulations to be performed.
We achieved this by a slight modification of i) and ii).

\begin{figure}
\hspace{0cm}
\vspace{-0.0cm}

\centerline{
\psfig{file=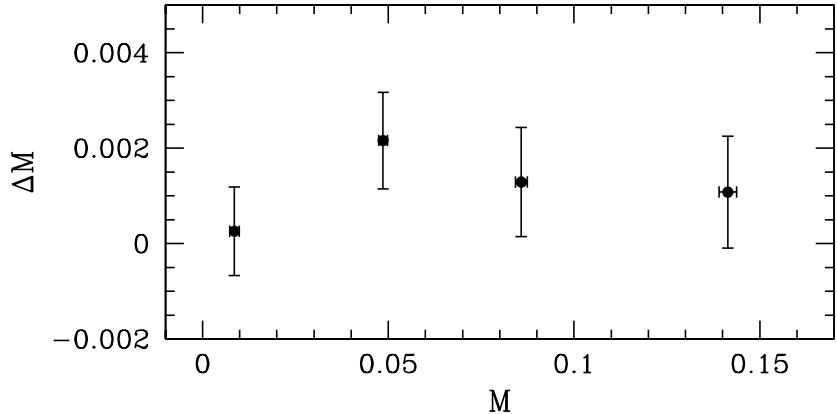,width=12cm}
}
\vspace{-0.0cm}
\caption{Mass dependence of the lattice artifact $\dM$ at 
$\beta=6/g_0^2=5.4,\csw=1.7275$.
\label{f_dM_M}}
\end{figure}

First of all, it was already  found in the quenched approximation
that $\dM$ is a very slowly varying function of $\M$ \cite{impr:pap3}.
Only negligible errors (compared with the statistical ones) 
are introduced if one keeps $\M$ just close to zero, say $|a\M| < 0.03$, 
instead of exactly zero. Of course, we must be careful when generalizing
from the quenched approximation since there
the quark mass enters only through the quark propagator
(valence quark mass), while here the quark mass is present in
the fermion determinant as well (sea quark mass).
For one set of parameters
$\csw,g_0$, we have therefore verified that also in the full 
theory $\dM$ depends only weakly on $\M$.  This is shown 
in \fig{f_dM_M}. Note that $g_0$ is chosen relatively large,
in order not to be in the situation where quark loops are trivially 
suppressed. 
Apart from this test, we have verified 
for each pair of $\csw,g_0$ that the dependence of $\dM$ on
the valence quark mass at
fixed sea quark mass is much smaller than the statistical
uncertainty of $\dM$, even when the valence quark
mass is increased to values of order $0.05/a$. 
From here on we therefore use $\dM$ for 
$|a\M| < 0.03$ as estimates for $\dM$ at $\M = 0$
rather than performing several runs and interpolating to 
$\M = 0$.
Note that most of our data for $\M$, given in
\tab{t_dm}, are in fact much smaller than our bound $|a\M| = 0.03$. 
Apart from the test just mentioned, we do of course always 
have valence and sea quarks
with the same mass.

\begin{table}[h]
 
\begin{center}
\caption{Results for the current quark mass $M$ and the 
lattice artifact $\Delta M$. 
}
\label{t_dm}

\vspace{0.5cm}
\begin{tabular}{|c|c|c|*{2}{r@{.}l|}}
\hline
 $\beta$ & $\hop$ &  $\csw$ & \multicolumn{2}{c|}{$aM$}  & \multicolumn{2}{c|}{$a\Delta M$} \\
\hline\hline
 12.0 & 0.12981 & 1.1329500& $-0$&$0102(2)$  & $-0$&$0007(2)$\\
 12.0 & 0.12981 & 1.1829500& $-0$&$0031(1)$  & $-0$&$0002(1)$ \\
 12.0 & 0.12981 & 1.2329500& $0$&$0038(1) $  & $ 0$&$0004(1)$  \\[0.6ex]
  9.6 & 0.13135 & 1.2211007& $-0$&$0031(2)$  & $ 0$&$0000(1)$ \\[0.6ex]
  7.4 & 0.13460 & 1.2155946& $-0$&$0008(5)$  & $ 0$&$0017(3)$   \\
  7.4 & 0.13396 & 1.2813360& $0$&$0037(3)$   & $-0$&$0001(3)$ \\
  7.4 & 0.13340 & 1.3445066& $0$&$0050(3)$   & $-0$&$0002(4)$  \\
  7.4 & 0.13245 & 1.4785602& $-0$&$0002(5)$  & $-0$&$0022(4)$   \\[0.6ex]
  6.8 & 0.13430 & 1.4251143& $0$&$0014(4)$   & $ 0$&$0000(3)$   \\[0.6ex]
  6.3 & 0.13500 & 1.5253469& $0$&$0013(6)$   & $-0$&$0004(4)$ \\[0.6ex]
  6.0 & 0.13910 & 1.2659000& $0$&$0087(7)$   & $0$&$0018(7) $  \\
  6.0 & 0.13640 & 1.5159000& $0$&$0025(7)$   & $-0$&$0002(6)$    \\
  6.0 & 0.13330 & 1.7659000& $0$&$0108(5)$   & $-0$&$0014(6)$  \\[0.6ex]
  5.7 & 0.14130 & 1.2798947& $0$&$005(1)$    & $0$&$0055(9) $   \\
  5.7 & 0.13770 & 1.5569030& $0$&$004(1)$    & $0$&$0007(7) $ \\
  5.7 & 0.13410 & 1.8339110& $0$&$0045(6)$   & $-0$&$0016(5)$ \\[0.6ex]
  5.4 & 0.14360 & 1.3571728& $0$&$023(3)$    & $0$&$004(4)  $\\
  5.4 & 0.13790 & 1.7275432& $0$&$009(1)$    & $0$&$0003(9) $\\
  5.4 & 0.13250 & 2.0979135& $0$&$007(2)$    & $-0$&$0016(8)$ \\[0.6ex]
  5.2 & 0.13300 & 2.0200000& $0$&$123(4)$    & $-0$&$0006(9)$  \\
\hline
\end{tabular}
\end{center}
\end{table}

Next, let us discuss step ii). In particular, we want to show that it
is not really necessary to perform calculations for several values
of $\csw$ for {\em each} value of the bare coupling $g_0$.
Let us denote by $\csw^{\rm impr}(g_0)$, the desired value
of $\csw$ for which the improvement condition is satisfied.
For $\csw$ close to $\csw^{\rm impr}(g_0)$,
$\dM$ will depend linearly on $\csw$ and 
the lattice artifact may be written as 
$\dM-\dM^{(0)} = \omega \cdot (\csw-\csw^{\rm impr})$, with a slope
$\omega(g_0)$ dependent on the gauge coupling. The numerical procedure
adopted in \cite{impr:pap3} consists of fitting $\dM$ to this linear
dependence separately 
for each value of the bare coupling $g_0$. 
To improve on this, we may use the fact that
the slope
$\omega(g_0)$  is expected to be a smooth function of $g_0$. 
Indeed, the numerical
values of $\omega$ as determined in the quenched approximation 
show that $\omega(g_0)$ can well be described by a linear behavior
in $g_0^2$,  
\bes
 \omega(g_0)=-0.015\cdot(1+\omega_1 g_0^2)\; ,
\ees
with a value of $\omega_1$ small such that $\omega(g_0)$ 
does not differ
much from the tree-level value $\omega(0)=-0.015$. This holds also 
for our results in full QCD as may be inferred from \tab{t_dm}.
It is therefore not necessary to determine $\omega$ for each value
of $g_0$ separately. Instead, we use its smoothness
to parametrize $\omega$
by an effective first order dependence on $g_0^2$, 
and perform one global fit to all our data of $\dM$ of the form
\begin{equation}
a\dM - 0.000277 = \omega(g_0)\cdot (\csw-\csw^{\rm impr}(g_0))  
\, .
\end{equation}
The fit parameters here are $\omega_1$ and the  desired values
$\csw^{\rm impr}(g_0^{(j)})$ at the different points $g_0^{(j)}$
where we have data.
As an aside we remark that the fit to the $\nf=2$ data,
$\omega = -0.015 \cdot(1 -0.33 g_0^2)$,  describes 
the slopes $\omega$ also for $\nf=0$.

\begin{figure}[htb]
\hspace{0cm}
\vspace{-0.0cm}

\centerline{
\psfig{file=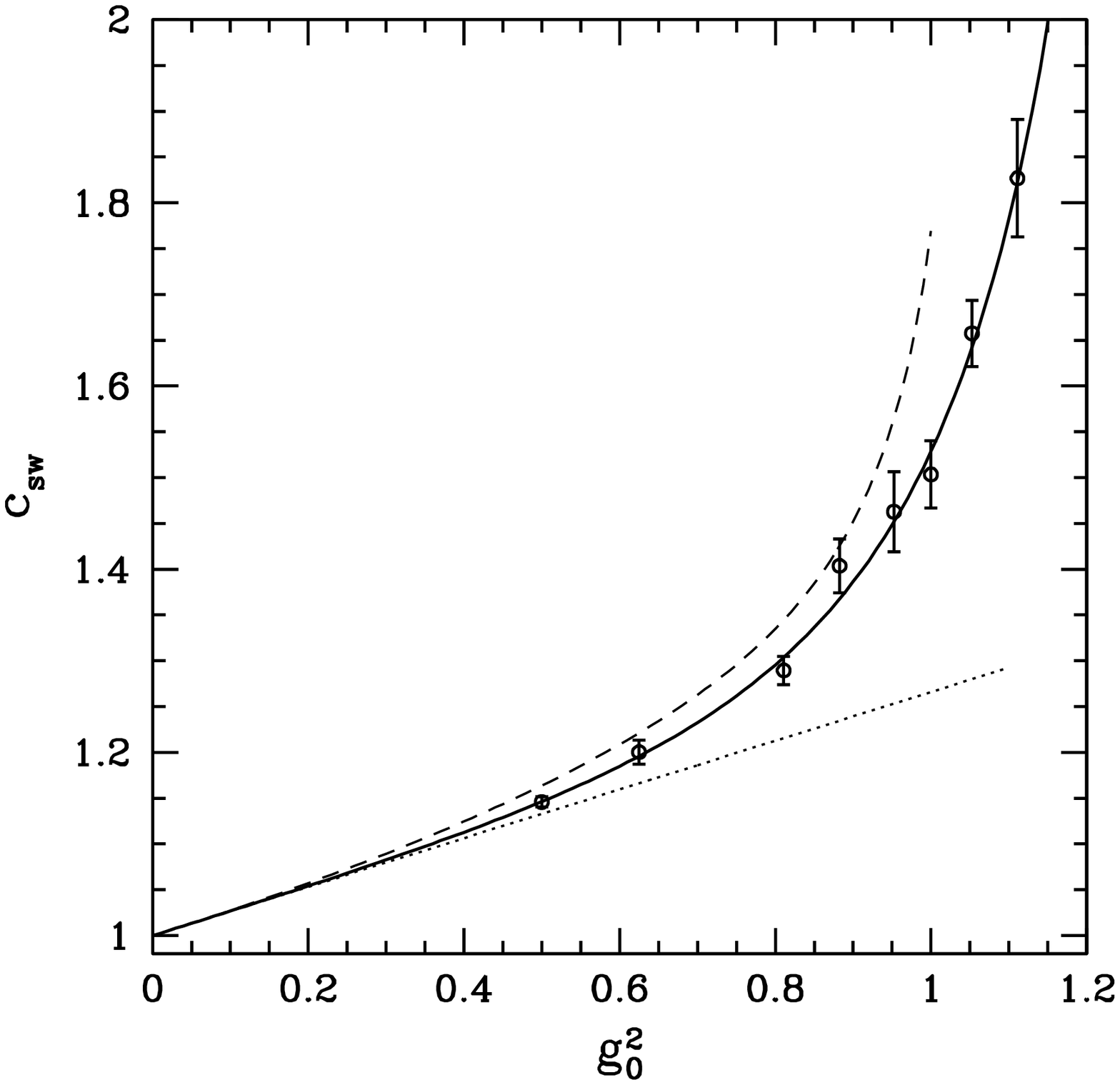,width=11.5cm}
}
\vspace{-0.0cm}
\caption{Non-perturbatively determined improvement coefficient $\csw$
for $\nf=2$. The dotted line shows first order perturbation theory 
\protect\cite{impr:Wohlert,impr:pap2} and the dashed curve is the result for $\nf=0$
\protect\cite{impr:pap3}.
\label{f_csw}}
\end{figure}
In this way, we obtain the values $\csw^{\rm impr}(g_0^{(j)})$ that 
satisfy our 
improvement condition. They are shown as data points in \fig{f_csw}.
In the whole range of $g_0$, they are well parametrized by 
\begin{equation} \label{pade} 
\csw=  
{1-0.454g_0^2-0.175g_0^4+0.012g_0^6+0.045g_0^8 \over
              1-0.720g_0^2} .
\end{equation} 
This representation, shown by the  full curve 
in \fig{f_csw},
is the main result of our work. It should be taken as a definition
of the improved action for future work. In this way it is guaranteed
that observables in the improved theory are smooth functions of the bare 
coupling and extrapolations to the continuum limit can be performed.

As in the quenched approximation, 
the $\nf=2$ result is well approximated by perturbation theory (\eq{e_csw1})
for small couplings, say $g_0^2 \leq0.5$. For larger 
couplings it grows quickly, although not quite as steeply
as in the quenched approximation (the dashed 
curve).

An important issue is the question for which range of couplings  
\eq{pade} is applicable. 
A priori it is to be trusted only for $\beta \ge 5.4$, 
where $\csw$ was computed by the numerical simulations. 
Extrapolations far out of this range are dangerous. 
We did, however, investigate whether it is justified
to use  \eq{pade} at somewhat smaller $\beta$. 
Unfortunately, already
for $\beta \approx 5.2$ the numerical simulations 
close to $M=0$ turned out to be too time consuming for our 
256 node APE-100 computer providing 6Gflop/s (sustained).
What helps again is that $\dM$ hardly depends on $M$. Therefore we expect to
find a small value of $\dM$  also at larger values of $\M$, say $|a\M | < 0.15$ (see 
\fig{f_dM_M}), if
the action is properly improved. Our calculation
at $\beta=5.2$ and $a\M \approx 0.12$ yielded
$a\dM=-0.0006(9)$, indicating that our improvement
condition is indeed satisfied for $\csw$ as given
by \eq{pade} for $\beta$ as low as $\beta=5.2$.
However, this calculation also revealed that higher order
lattice artifacts rapidly become stronger 
when $\beta$ is taken  below $\beta=5.4$. We will return
to this issue in \sect{s_asq}.

\section{Estimate of $\hopc$}

\begin{figure}
\hspace{0cm}
\vspace{-0.0cm}

\centerline{
\psfig{file=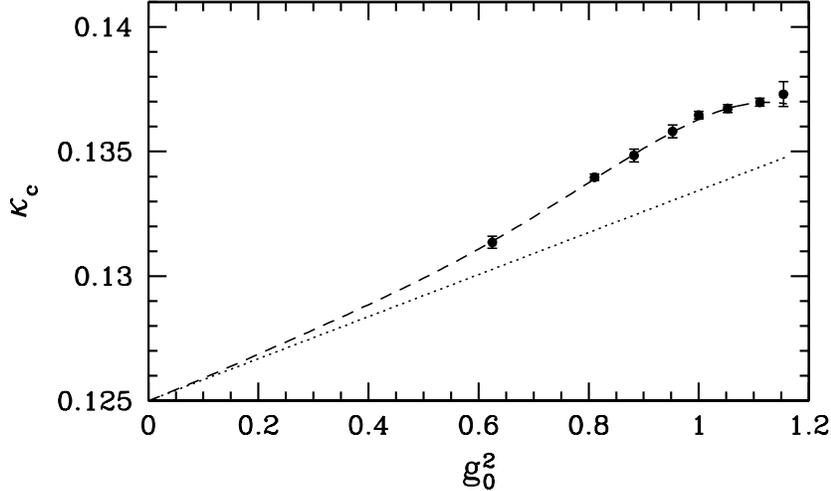,width=12cm}
}
\vspace{-0.0cm}
\caption{The critical line in the improved theory. 
The dashed curve gives the polynomial 
approximation to the non-perturbative result and the dotted line
indicates first order perturbation theory.
\label{f_kc}}
\end{figure}

For future applications of the improved action, it is useful to
roughly know the position of the critical line 
\bes
 \hop = \hopc(g_0),\quad  \hop \equiv \frac{1}{2(am_0+4)}, \label{e_crit_line}
\ees
which is defined by the vanishing of the current quark mass. We will
give an estimate for $\hopc(g_0)$ in this section.
The critical line \eq{e_crit_line}
has an intrinsic uncertainty, since the position where the current quark mass
vanishes depends on the very definition of the current quark mass. 
The uncertainty in the current quark mass is $\Oasq$, 
translating into an $\rmO(a^3)$ uncertainty in
$\hopc$. From Fig.~9 in ref.~\cite{impr:pap3}, we estimate 
that the values
of $\hopc$ that one determines on a lattice with $L/a=8$, might differ
from  $\hopc$ determined on larger lattices by as much
as $2 \times 10^{-4}$. This has to be kept in mind as an important limitation of the 
present determination of $\hopc$.

Our basis for an estimate 
of $\hopc$ are the numerical data for $M$ at a number of values of the 
parameters $\hop$, $g_0$, $\csw$. 
For $\beta \ge 5.4$ the values of $aM$ are rather small, see \tab{t_dm}. One 
can convince
oneself easily that it is justified in this case to
use the 1-loop relations \cite{impr:pap5,impr:stefan_notes}:
\bes
 \M &=&  Z_{\rm m} \mq \, (1+  b a \mq), 
 \quad a \mq = {1 \over 2 \hop} - {1 \over 2 \hopc} \\
 b &=& -1/2 - 0.0962 \cdot g_0^2 , \quad  Z_{\rm m}= 1 + 0.0905 \cdot g_0^2 ,
\ees
to determine $\hopc$ from $\hop$, $a\M$.
The uncertainties due to left out higher order terms in the above equations
may be neglected 
compared to the statistical uncertainties in $\M$, since 
$\hopc$ is close to $\hop$ in any case.  
As mentioned earlier, for $\beta=5.4, \csw=1.7275$,
we have a series of different values of $\hop$.
We have investigated whether the 1-loop relations describe 
$\M(\hop)$ in this case as well.  Up to $a\M \approx 0.14$
and within an error
margin of about 5\%,  this is 
shown to be the case by our data.
Therefore we also included the point
$\beta=5.2$, $a\M \approx 0.12$ 
in the analysis -- despite the relatively large
mass of that point.
The statistical uncertainties in $a\M$ are 
then translated into uncertainties in $\hopc$.

Next, we need to interpolate $\hopc$ in $\csw$ to the proper
values given by \eq{pade}. These are denoted 
by $\csw^{\rm impr}(g_0)$ below. In this interpolation, we use again that 
the slope of $\hopc$
as a function of $\csw$ is a smooth function of $g_0$. 
We fit all values of $\hopc$ to
\bes
\hopc &=& \hopc^{\rm impr}(g_0) + k(g_0) \cdot (\csw -\csw^{\rm impr}(g_0)) ,
\label{e_kcfit} \\
k(g_0)&=& \sum_{i=1}^3 k_i (g_0)^{2i}.
\ees
Here, $k_1$ is set to its perturbative value of $-0.053/8$ \cite{impr:QCDSF_lett},
while $k_2,k_3$ and the desired values $\hopc^{\rm impr}(g_0^{(j)})$ are
fit parameters. \Eq{e_kcfit} fits all values of $\hopc$ within an
error margin of $2 \cdot 10^{-4}$, which is also roughly the statistical 
accuracy of $\hopc$.  

The results of the fit, $\hopc^{\rm impr}(g_0)$, are shown as data points
in \fig{f_kc}. They are well approximated by the polynomial (dashed curve),
\bes
  \hopc&=&1/8+\hopc^{(1)}g_0^2+0.0085g_0^4 -0.0272 g_0^6+ 0.0420 g_0^8 
        -0.0204g_0^{10} \; , 
  \label{e_fitkc}      
         \\
         && \hopc^{(1)}=0.008439857 \; .   
\ees
The critical line is never very far from the 1-loop result $\hopc=1/8+\hopc^{(1)}g_0^2$
\cite{pert:gabrielli,impr:Wohlert,impr:pap2}.

We emphasize once more that we regard \eq{e_fitkc} 
as a first estimate and expect only a rather crude 
precision (statistical + systematic) of order $\pm 3 \cdot 10^{-4}$.

\section{$\Oasq$ effects after improvement \label{s_asq}}

\begin{figure}[ht]
\hspace{0cm}
\vspace{-0.0cm}

\centerline{
\psfig{file=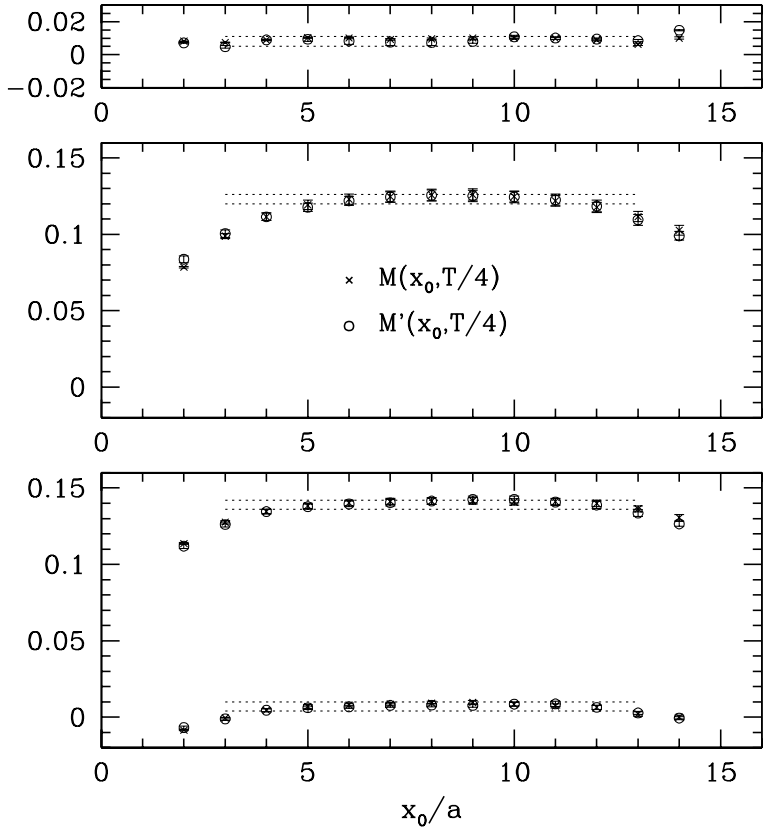,width=12cm}
}
\vspace{-0.0cm}
\caption{$\M$ and $\M'$ for $\beta=5.4,\csw=1.7275$ (bottom part of the figure)
and $\beta=5.2$, $\csw$ as given by \eq{pade} (middle).
The very top section is for $\beta=6.0, \csw=1.8659, \nf=0$. 
The time extent of the 
lattice is $T = 16a$.
\label{f_asq}}
\end{figure}

Once the improved action is known up to ${\rm O}(a)$, 
a new question arises immediately: how large are higher order lattice 
artifacts after improvement? In the quenched approximation this has
been investigated in the \SF and also for low-energy hadronic observables
\cite{impr:QCDSF_lett,impr:roma2_2,impr:qcdsf,lat97:hartmut,Edwards:1997nh}. 
Only rather small 
$\Oasq$ effects have been found.
In the full theory, such investigations will
still take some time. 
Since our observable $\M(x_0,T/4)$ should not depend on $x_0$, apart from
the $\Oasq$ effects, it might serve meanwhile as estimator of the
higher order lattice artifacts. 

We plot $a\M(x_0,T/4)$ and $a\M'(x_0,T/4)$ in the lower part of \fig{f_asq},
for $\beta=5.4$ and a value of $\csw$ where $\dM$ almost vanishes as is obvious
from the agreement between $\M$ and $\M'$.
Results for two values of $\hop$ are shown. They correspond to rather 
different $M$.
At both values of $M$ 
we observe that the variations of $a\M(x_0,T/4)$ and $a\M'(x_0,T/4)$ are within a
corridor of $\pm 0.003$ as long as $4a \le x_0 \le T-4a$.
As one takes $x_0$ closer to the boundaries, the values of $\M$ drop
significantly below the values of $\M$ belonging to the corridor. 

The middle part of the figure shows the situation for $\beta=5.2$.
Here, $x_0=4a$ is already outside of a corridor 
of the same size and $x_0=3a$
is quite far below. This indicates that at $\beta=5.2$ the $\Oasq$ lattice
artifacts are already quite significant.  Of course, their impact on
quantities such as the hadron spectrum remains to be investigated in detail.
Nevertheless, our analysis indicates that for values of $\beta$ 
below $\beta=5.2$, $\Oa$ improvement
ceases to be very useful.

The above conclusion is further strengthened by a comparison 
with the quenched approximation (top part of the figure) where -- at a
relatively large lattice spacing of $a\approx0.1\fm$ -- 
the $x_0$-dependence is hardly visible.

In interpreting the lattice artifacts in $\M(x_0,T/4)$,
one must be careful about the following point. Close to the boundaries,
$x_0=0$ and $x_0=T$, the spectral decomposition of the correlation functions,
$f$, $f'$, receives noticeable contributions also from intermediate states
with energies of the order of the cutoff. 
In such a kinematical regime, on-shell improvement is not applicable. 
We should therefore not
put too much weight on the behavior very close to the boundaries.
(For this reason we are not showing the points $x_0=a$ and $x_0=T-a$ in
the figures.) 
Even with this reservation in mind, the difference between 
the quenched approximation at $a\approx 0.1\fm$ and $\nf=2$ QCD at
$\beta=5.2$ is striking. 

We further note that preliminary results of the UKQCD collaboration
\cite{UKQCD:unpubl} indicate that the lattice spacing is 
 larger than $0.1\fm$ at $\beta=5.2,\nf=2$. 
This is in line with the considerable
size of lattice artifacts
visible in \fig{f_asq}.

\section{The Hybrid Monte Carlo simulation \label{s_HMC}}

In this section we want to present a number of aspects of the
simulations we have performed. All numerical results quoted in
this paper have been obtained by using the Hybrid Monte Carlo (HMC) algorithm 
\cite{Duane:1987de}.
One simulation (at $\beta=6.8$) was repeated with the Polynomial
Hybrid Monte Carlo algorithm \cite{Frezzotti:1997hc,Frezzotti:1997ym}, 
and completely
consistent values were found for all observables compared \cite{Frezzotti:1998_prep}. 

Our particular implementation of the HMC algorithm using even/odd
preconditioning \cite{DeGrand:1990dk} and including
the improvement term of \eq{e_dimpr} is described in detail in \cite{Jansen:1997yt}. 
Throughout the simulations we used the higher order 
leap-frog integrator
suggested in \cite{Sexton:1992nu} to integrate the equations
of motion to a trajectory length of one,
implementing eq.~(6.7) of ref.~\cite{Sexton:1992nu} with $n=4$.

The program was run on the Alenia Quadrics (APE) massively
parallel machine with 256 nodes. We decided to distribute
our lattice of size $16\cdot 8^3$ on these machines in such a way that
we ran $\nrep=32$ replica in parallel. 
These replica are independent copies of the lattice
for identical choices of all parameters (apart from random numbers);
we end up with $\nrep$ statistically independent simulations
for each set of parameters.

When starting our simulations, we tried to keep the acceptance rate
to about $90\%$. However, 
it happened rather frequently that,
 despite this relatively large acceptance rate, 
in one of the replica 
a considerable number of trajectories in a row were rejected,
inducing substantial autocorrelation times. 
Our solution to 
overcome this problem, was to perform every $\nsafe$ trajectories one
with a much smaller step size, which we call the 
``safety step''.  In principle, $\nsafe$ as well as the corresponding 
step size can be drawn randomly from arbitrary distributions,
while still preserving detailed balance of the HMC algorithm.
The choice of $\nsafe$ and the step size must, however, not depend 
on the Monte Carlo history. 
For simplicity we chose fixed values for $\nsafe$ and the step size before 
each run, determining reasonable values from our experience 
gained in simulations at other $\beta$'s and also during the 
thermalization phase.

\begin{table}[t]
 
\begin{center}
\caption{The parameters of the runs (we typically had $\nsafe=6$). 
We denote by
$\delta\tau$ the step size of the leap-frog integrator, 
giving in brackets the value of the safety step size. 
$\pacc$ denotes the acceptance rate and 
$N_{\rm CG}$ the average number of Conjugate Gradient 
iterations per inversion. 
}
\label{t_runparameters}

\vspace{0.5cm}
\begin{tabular}{|r|c|c|l|c|l|c|}
\hline
 $\beta$~ & $\hop$ &  $\csw$ & ~~$\delta\tau(\delta\tau_{\rm save})$
         & $\statis$ & ~~~$ \pacc $ & $\ncg$ \\
\hline\hline
 12.0 & 0.12981 & 1.1329500& 0.040       & 1600 & 0.976(3) & 129.5(0.5)\\
 12.0 & 0.12981 & 1.1829500& 0.040       & 3264 & 0.981(2) & 130.0(0.4)\\
 12.0 & 0.12981 & 1.2329500& 0.040       & 2080 & 0.978(3) & 130.8(0.6)\\
  9.6 & 0.13135 & 1.2211007& 0.033       & 2304 & 0.985(3) & 137.0(0.6)\\
  7.4 & 0.13245 & 1.4785602& 0.066(0.04) & 1280 & 0.890(10)& 138.3(1.3) \\
  7.4 & 0.13460 & 1.2155946& 0.066(0.04) & 1632 & 0.902(10)& 139.4(3.0)  \\
  7.4 & 0.13396 & 1.2813360& 0.071(0.04) & 3040 & 0.887(10)& 133.8(0.2) \\
  7.4 & 0.13340 & 1.3445066& 0.033       & 1856 & 0.987(1) & 154.9(0.7)\\
  6.8 & 0.13430 & 1.4251143& 0.066(0.025)& 3104 & 0.859(12)& 145.9(3.0) \\
  6.3 & 0.13500 & 1.5253469& 0.047(0.04) & 1536 & 0.947(8) & 165.2(1.3)\\
  6.0 & 0.13330 & 1.7659000& 0.033(0.02) & 2080 & 0.958(6) & 172.1(2.5)\\
  6.0 & 0.13640 & 1.5159000& 0.033(0.02) & 2336 & 0.888(17)& 186.2(4.5)  \\
  6.0 & 0.13910 & 1.2659000& 0.033(0.02) & 1856 & 0.966(3) & 160.2(3.0) \\
  5.7 & 0.14130 & 1.2798947& 0.037(0.033)& 1600 & 0.963(6) & 168.7(3.0) \\
  5.7 & 0.13410 & 1.8339110& 0.040(0.033)& 2528 & 0.920(10)& 192.3(2.5) \\
  5.7 & 0.13770 & 1.5569030& 0.037(0.033)& 2720 & 0.963(5) & 179.5(2.0)\\
  5.4 & 0.13790 & 1.7275432& 0.033(0.027)& 5120 & 0.948(8) & 208.7(2.5)\\
  5.4 & 0.14360 & 1.3571728& 0.027(0.01) & 3200 & 0.974(5) & 205.6(1.4)\\
  5.4 & 0.13250 & 2.0979135& 0.027(0.01) & 5760 & 0.958(7) & 216.8(1.0)\\
  5.2 & 0.13300 & 2.0200000& 0.030(0.025)& 3072 & 0.964(2) & 163.3(1.1) \\ 
\hline
\end{tabular}
\end{center}
\end{table}
 
We give in \tab{t_runparameters} the most relevant parameters of our 
simulations.
By $\ntraj$ we denote the number of trajectories 
that   
equals the number of measurements { per replicum}.
The strategy in performing the simulations was to start at a
high value of $\beta=12$ and then to decrease $\beta$ successively to the 
values
given in \tab{t_runparameters}. 
We found, when changing from one $\beta$ to the next smaller value, 
that only a short thermalization time of the order of ten trajectories
was required. 

\subsection{Dynamics of the HMC -- cost of a simulation}

To achieve our physics goal, we performed quite
a large number of simulations with different
values of the parameters $\kappa$, $\beta$ and $\csw$. 
Although this was not our main
objective, it is natural to also attempt to learn something about
the dynamics of the algorithm. Before doing so, let us recall
the special physical situation where these simulations were performed.
We kept $L/a$ and $T/a$ fixed, and stayed close
to the critical line. 
In contrast to the situation in infinite volume, 
the infrared cutoff is given by $T$ and the quark mass, once small, 
plays only a minor role for the physics and 
hence also
for the dynamics of the HMC algorithm. For the latter, a relevant quantity is
the  condition number,
\begin{equation} \label{condition_number}
k =  {\lambda_{\rm max} \over \lambda_{\rm min}} \; ,
\end{equation}
where $\lambda_{\rm min}$ ($\lambda_{\rm max}$) are the
lowest (largest) eigenvalues 
of the preconditioned matrix $\hat{Q}^2$, see \cite{Jansen:1997yt}.
This quantity was computed in all simulations, using the method of
ref. \cite{Bunk:eigen,Kalkreuter:1996mm}. 

For quark mass zero and with \SF boundary conditions, 
$k$ has a value of order $(T/a)^2$ for
free fermions~\cite{SF:stefan1}. 
In the interacting theory this 
is modified by terms of order $g_0^2$. 
For the MC dynamics, the inverse square root of the 
condition number is expected to roughly take over the role that
the quark mass plays for infinite volume simulations.

\begin{figure}[ht]
\hspace{0cm}
\vspace{-0.0cm}

\centerline{
\psfig{file=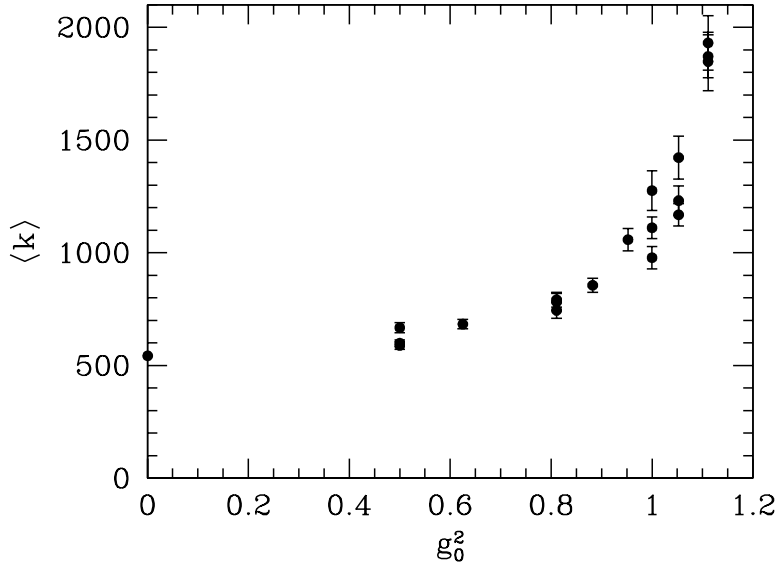,width=10cm}
}
\vspace{-0.0cm}
\caption{The expectation value of the condition number, 
eq.(\ref{condition_number}). 
When more than one value is plotted at a given value of $g_0^2$ 
they correspond to different values of $c_{\rm sw}$.
The tree-level value is shown at $g_0^2=0$.}
\label{fig_condition}
\end{figure}

In \fig{fig_condition} we see a strong rise
of $\langle k \rangle$ 
when $g_0^2$ is increased\footnote{The average is the usual ensemble average.}. For the largest values 
of the bare coupling, it is a factor three larger than 
the tree-level value at $g_0^2=0$. Of course, this is immediately
reflected in a considerable rise in the number of CG iterations needed 
for the various ``inversions'' of $\hat{Q}^2$. In fact, in our particular
implementation of running $\nrep$ simulations in parallel on a 
SIMD machine, there is an additional important overhead: the inversions
have to run until the ``slowest'' replicum has converged. 
Therefore the number of CG iterations depends
on the maximum of $k$ over the number of replica, denoted 
by $k_{\rm max}$. 
Since also the relative variance of $k$, defined by
$\langle k^2 \rangle / \langle k \rangle^2 -1$ grows from
around $0.1$ at $\beta=12$ to $0.5$ at $\beta=5.4$, we find
that $\langle k_{\rm max} \rangle / \langle k \rangle$ can be as 
large as $\langle k_{\rm max} \rangle / \langle k \rangle \approx 4$.
We have not exactly quantified the corresponding loss in speed of the
HMC as a function of $\nrep$, but expect that this may result in an
overhead of 20-40\% for $\nrep=32$~\footnote{
We remark that the condition number also develops a significant
dependence on the quark mass when $\beta$ becomes small and/or
large quark masses are considered. 
For example at $\beta = 5.4$, $\csw = 1.7275$ 
we find $\langle k \rangle = 1871(95)$ at $\M = 0.009(1)$ whereas
$\langle k \rangle = 800(14)$ at $\M = 0.086(2)$.}.

\begin{figure}[ht]
\hspace{0cm}
\vspace{-0.0cm}

\centerline{
\psfig{file=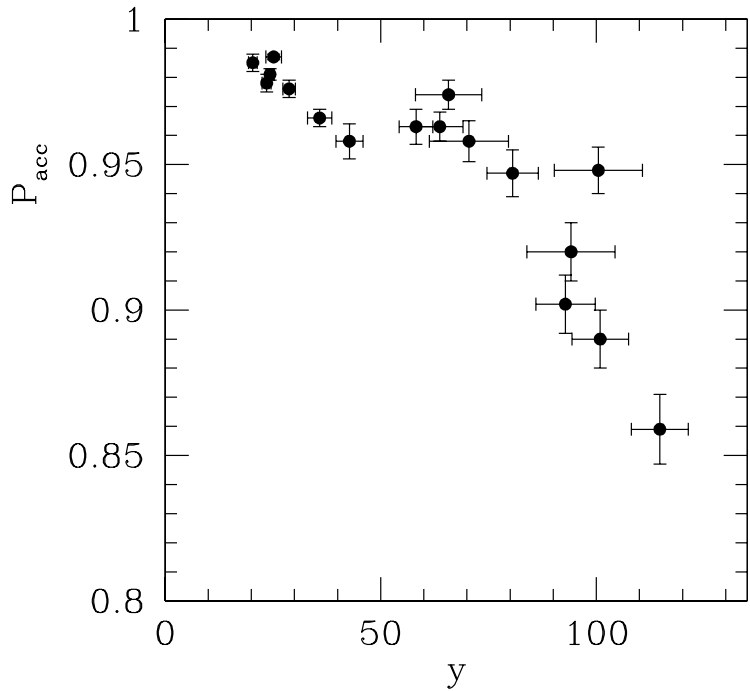,width=10cm}
}
\vspace{-0.0cm}
\caption{The acceptance rate as a function of 
$y=\delta\tau^2 \langle k^{3/2}\rangle$. Only data with $|aM|<0.03$ are 
included.}
\label{f_pacc}
\end{figure}
In addition to its direct relation to $\ncg$, the 
condition number also is an important
parameter, which has an influence on how large
$\dtau$ may be chosen for a desired value of the acceptance
$\pacc$.   One may argue  \cite{Gupta:1989kx,Gupta:1990ka} 
that  -- for our leap-frog
integrator -- $\pacc$ is approximately a universal function of the combination
$y=(\dtau)^2 \langle k^{3/2} \rangle $. We observe a rough consistency
with such a scaling law (cf. \fig{f_pacc}).

As already mentioned earlier,
the combination of the above effects leads to a large increase
in the cost of the simulations when $g_0^2$ is increased. 
In the following section we will see that autocorrelations
(mildly) enhance this effect further. This was one of the reasons why
we did not decrease $\beta$ below $\beta=5.4$, where we invested
already of the order of a 
month of CPU-time on our 6~Gflop/s (sustained) machine.
Given that the length of our lattice might be around $2\fm$ or larger
for $\beta<5.4$, it is in fact not 
a great surprise that simulations with very light quarks become
difficult in this regime. 

\subsection{Error analysis and autocorrelation times \label{s_errors}}

We used two different methods to obtain the errors of our
observables. The first one is to 
average observables over all measurements done
within one replicum. One is then left with $\nrep=32$ statistically
independent measurements. The errors,
computed by jack-knife, then 
have no systematic uncertainty due to autocorrelations
but they are only subject to a relative statistical uncertainty of
$(2 \nrep)^{-1/2} = 12.5\%$. 

\begin{figure}[ht]
\hspace{0cm}
\vspace{-0.0cm}

\centerline{
\psfig{file=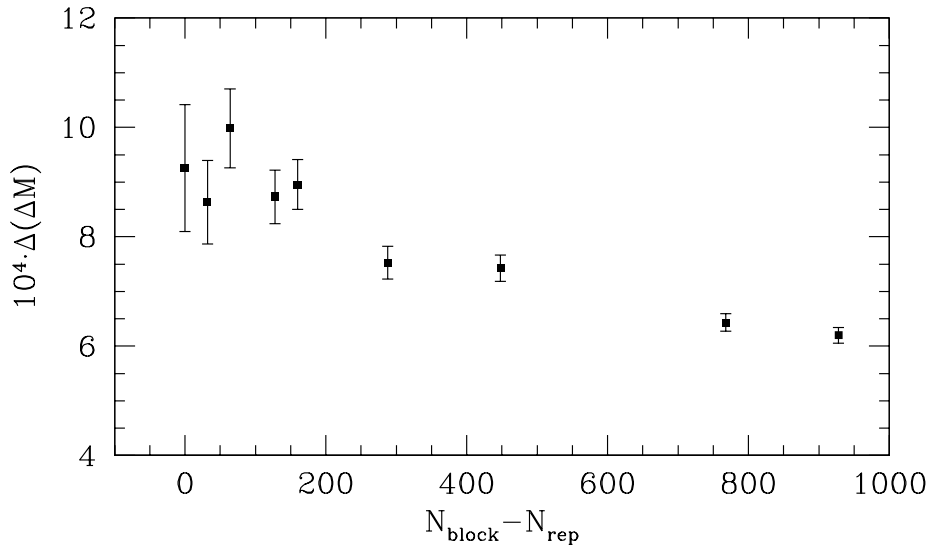,width=12cm}
}
\vspace{-0.0cm}
\caption{The error of the lattice artifact as a function of the 
         number of blocks $\nbl=\nrep\ntraj/\lbl$. 
         The naive error is $\Delta_{\rm naive}=4.5 \cdot
         10^{-4}$.
          The data are shown for parameters
         $\beta=5.4$, $\kappa=0.1379$, $c_{\rm sw}=1.7275432$.}
\label{f_blockerror}
\end{figure}

Alternatively we performed  a jack-knife
procedure combined with the following blocking analysis.
We generated an ensemble of
blocked measurements by averaging (for each replicum) 
subsequent measurements
over blocks of length $\lbl$. The blocked 
measurements of different replica were joined to form one
common sample (with $\nbl=\nrep\ntraj/\lbl$ blocks) from which 
we then computed the
jack-knife error $\Delta(O)$ of the observable $O$. 
For large statistics, these errors will have a negligible statistical
uncertainty, but still suffer from systematic corrections  
due to autocorrelations.
For autocorrelation times $\tau$, which are
small with respect to $\lbl$,  the {\em systematic} effect 
due to autocorrelations is 
proportional to $\tau /\lbl$,
while
the relative {\em statistical} uncertainty of this error estimate
is approximately given by  $(2 \nbl)^{-1/2}$.
A typical situation is shown in \fig{f_blockerror} for the error
of the lattice artifact $\dM$.
Since the autocorrelations for this quantity are not very large
(see below),
the error estimates converge quite well as $\nbl \to \nrep$\footnote{
We then have $\lbl=\ntraj$ and are back to the first method,
where there is no systematic uncertainty of the error estimates.
A little thought reveals that in general
one expects an approximately linear behavior in 
$\nbl-\nrep$ as suggested by \fig{f_blockerror}.}.
The most precise
estimate of the true error of $\dM$ would probably be given 
by an extrapolation to $\nbl=\nrep$ from larger values of $\nbl$. 
On the other hand,
there is a systematic (and subjective) bias in such an extrapolation
and we prefer to quote the unbiased error estimate discussed before, which 
has a satisfactory 
statistical precision of 
$12.5\%$ anyhow.

\begin{figure}[ht]
\hspace{0cm}
\vspace{-0.0cm}

\centerline{
\psfig{file=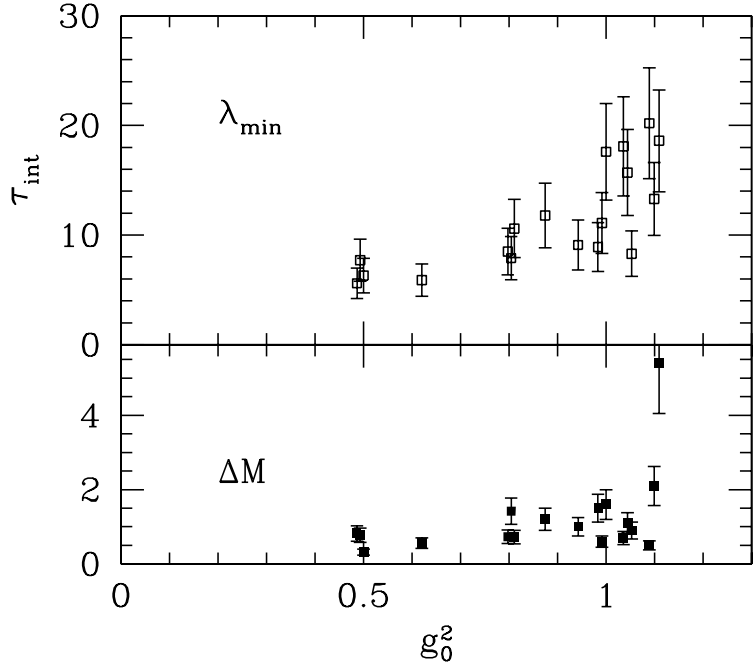,width=10cm}
}
\vspace{-0.0cm}
\caption{The integrated autocorrelation times of the lowest
         eigenvalue $\lambda_{\rm min}(\hat{Q}^2)$ and the
         lattice artifact $\Delta M$ as a function of the
         bare coupling. When more than one value is plotted 
         at a given value of $g_0^2$, 
         they correspond to different values of $c_{\rm sw}$.
}
\label{autocorrelations}
\end{figure}

Our definition of the integrated autocorrelation time $\tauint$
of an observable $O$ is 
\begin{equation} \label{autoint}
\tau_{\rm int}(O) =\frac{1}{2} \left({\Delta(O) \over 
                                      \Delta_{\rm naive}(O)}\right)^2\; ,
\end{equation}
where
$\Delta_{\rm naive}$ is the naive error (computed with $\lbl=1$).
Estimating the true error as described above, i.e. having 
$\lbl=\ntraj$, this quantity has 
a statistical uncertainty of approximately
$\Delta(\tau_{\rm int}) = \sqrt{2/\nrep} \cdot \tau_{\rm int}$.
We show in \fig{autocorrelations} the integrated autocorrelation 
times for the lowest eigenvalue $\lambda_{\min}(\hat{Q}^2)$ and for
our main observable $\dM$. 
The figure indicates that the autocorrelation times 
increase with growing bare coupling. 
Furthermore, the integrated autocorrelation times depend 
strongly on the observable:
for the lattice artifact $\Delta M$, $\tauint$ is always much below
the one for $\lambda_{\min}(\hat{Q}^2)$. 
We note that, on the contrary, the integrated
autocorrelation times for the correlation functions $f_A$ and $f_P$,
from which $\Delta M$ is derived, 
are similar to the ones for $\lambda_{\min}(\hat{Q}^2)$.
Even larger autocorrelation times for other observables  
have been observed after cooling. As discussed in the following subsection,
it is, however, very unlikely that they invalidate our error estimate 
for $\dM$.

\subsection{Metastable states}

\begin{figure}[ht]
\hspace{0cm}
\vspace{-0.0cm}

\centerline{
\psfig{file=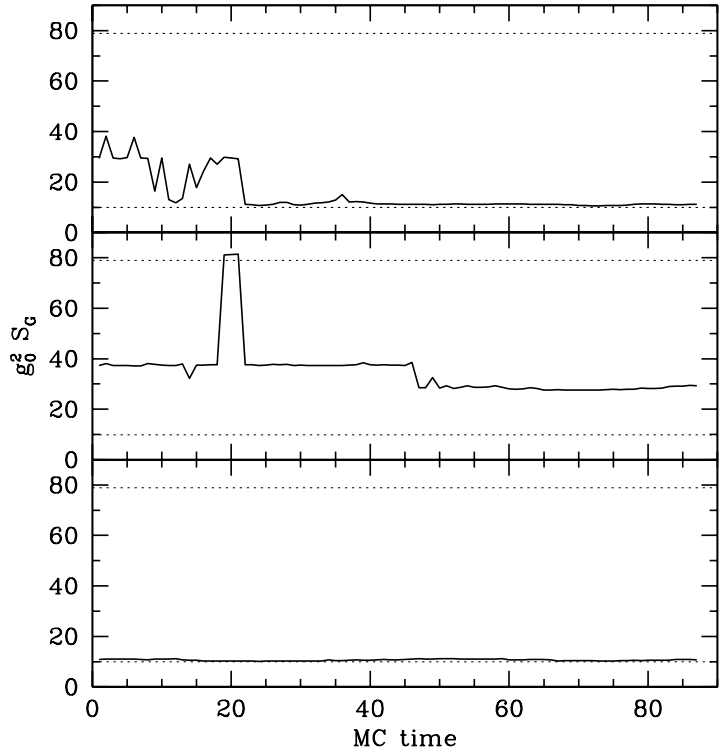,width=9.5cm}
}
\vspace{-0.0cm}
\caption{Part of the HMC history of the gauge field action $\Sg$ after
cooling. We show three of our 32 replica. Simulation parameters are
$\beta=5.4, \hop=0.1325, \csw=2.0979$. The bounds (\ref{e_act_bound}) and 
(\ref{e_act_inst}) are shown as dotted lines. 
\label{f_hist}}
\end{figure}
As an additional interesting observable we have also monitored
the renormalized coupling constructed as described in \cite{alpha:SU3}
(with the derivative with respect to the background gauge field 
acting only on the pure gauge action).
While for the larger values of $\beta$, this quantity showed
autocorrelation times of the same order as $\tauint(\dM)$, we observed
a considerable rise in the integrated autocorrelation times 
eq.(\ref{autoint})
for $\beta \leq 5.4$. We were worried that this behavior 
as well as the large values of $\tauint(\lambda_{\rm min})$
are due to difficulties of the HMC algorithm to sample different
topological sectors as they have been observed before
in large volume simulations \cite{hmc:topo1} (see also 
\cite{hmc:topo2}). 

In order to investigate this possibility, we 
also examined the gauge fields after performing a number of
cooling iterations \cite{topo:cool}, 
computing among other observables
the gauge-field action and the topological charge $Q$
(``naive definition'', see \cite{topo:naive_charge}).
For our abelian background field, the gauge field action 
satisfies the bounds \cite{SF:LNWW} (a small $\Oa$ correction
is neglected here)
\bes
 g_0^2 \Sg & \geq & \pi^2, \quad Q=0 \label{e_act_bound},\\
 g_0^2 \Sg & \geq & 8\pi^2 |Q| \label{e_act_inst}.
\ees 
We note that 
\eq{e_act_inst} is derived for smooth fields in the continuum
Schr\"odinger functional, while \eq{e_act_bound} is valid also for rough fields. 

In all our runs where we performed 
 cooling, we observed only once a value of 
$Q$ different from zero. The short Monte Carlo (MC) time interval
where this happened 
is contained in the MC history shown in the 
middle part of \fig{f_hist}, where $g_0^2S_G \approx 80$. Exactly during the 
interval
where the action is above 
the limit \eq{e_act_inst} with $|Q|=1$, we also observe that
$Q$ 
has an  (approximately) unit value.
Given their rareness,
topological fluctuations appear not to be relevant
in our small volume simulations and there is an indication
that they are in fact not long-lived.
In particular, slow {\em topological} fluctuations are
clearly not the cause of the
large autocorrelation times observed in our simulations. 

Most of the time, the action (always after cooling in this section) is
close to the absolute minimum of \eq{e_act_bound}. However,
we also observed longer sections in the MC-history, where it
remains at other values (e.g. $g_0^2 \Sg  \approx 40$). 
These appear to correspond to 
non-trivial local minima of the action (with $Q=0$). Since 
these states are stable over 
several tens of trajectories, there is a mode
with a very large 
autocorrelation time in the HMC simulations. 

We now have to investigate whether our observable $\dM$ is affected 
by this large autocorrelation time, and the small $\tauint$
determined in the previous section is misleading. 
We therefore want to know the correlation of 
$\dM$ with these states. As a measure of such a
correlation we consider the linear 
correlation coefficient $\cor(\dM,g_0^2 \Sg)$. A standard definition of
the correlation 
coefficient of primary observables, $a,b$, 
(i.e. observables
that are given directly as ensemble averages) is 
$\cor(a,b)=\cov(a,b)/[\cov(a,a) \cov(b,b)]^{1/2}$,
where   
$\cov(a,b)=\langle\, (a-\langle a\rangle )
                     (b-\langle b\rangle )\,\rangle$.
Since we are mainly interested in $\dM$, a derived quantity (i.e. a
function of primary observables),
we generalize the above definition to 
\bes
  \cor(a,b) = \left\{ \sigma^2(a b) - \langle b\rangle^2\sigma^2(a) - 
                                      \langle a\rangle^2\sigma^2(b) \right\}
             / \left\{ 2 \langle a\rangle\langle b\rangle 
             \sigma(a)\sigma(b)\right\} \; ,
\ees
which can easily be used also for derived quantities, by computing
the variances $\sigma^2(a)$ by a jack-knife analysis.

We found very small correlation coefficients between 
$\Sg$ and all 
fermionic observables considered, where of course the fermionic 
observables  are the ones that enter the physics, i.e. they are 
computed without cooling. As an example we quote
$\cor(\dM,g_0^2\Sg)=0.02$ for 
$\beta=5.4, \csw=2.0979$. We conclude that the metastable
states that we observed are not a matter of major concern for our 
error analysis in the determination of $\csw$.

This corroborates our error analysis of \sect{s_errors}.

\section{Conclusions}

In this paper we have performed the first step to 
achieve a non-perturbatively $\Oa$ improved lattice
theory for Wilson fermions.  We have computed the improvement
coefficient $\csw$ as a function of $\beta=6/g_0^2$          
in a range of couplings $\beta \ge 5.2$. 
This range of couplings seems to cover values 
of lattice spacings where simulations for e.g. studying hadronic properties
can be performed. 
As our main result we consider 
the parametrization of \eq{pade}, which determines the non-perturbatively
improved 
action for $\nf =2$ dynamical flavors of Wilson fermions. 
A number of quantities such as the hadron spectrum can now be computed
with lattice artifacts starting only at order $a^2$ and, indeed,
such a programme has been initiated already \cite{UKQCD:unpubl}.
In order to achieve the same accuracy for hadronic matrix elements,
also the improvement and normalization of the
operators have to be determined along 
the lines exploited already in the quenched approximation 
\cite{impr:pap4,lat97:marco,impr:roma2_1} or following new suggestions
\cite{impr:roma1}.

Although after $\Oa$ improvement the linear lattice artifacts 
are cancelled,
higher order discretization errors will remain. 
Indeed, we found indications that for $\beta < 5.4$
these effects can become significant. 
It would be desirable to investigate these effects further
for additional physical observables. 

During the course of our small volume simulations,
we were also able to study the dynamical behavior of
the Hybrid Monte Carlo algorithm used throughout
this work. We found that the condition number 
of the fermion matrix rises with increasing 
coupling strength. The condition number
directly influences various ingredients of the
algorithm: changing $\beta$ from large values to $\beta=5.4$, 
we found that
the number of Conjugate Gradient iterations increased by a factor
  of about $1.6$, 
the step size had to be decreased by a factor of $2$ and 
the autocorrelation times increased by about a factor 
of $2$. 
All these effects add up to make simulations 
very expensive when $\beta$ is chosen to be $\beta \approx 5.2$ or
even smaller.\\[2.0ex] 
{\bf Acknowledgements}\\
This work is part of the ALPHA-collaboration research programme.
We thank DESY for allocating computer time on the APE/Quadrics
computers at DESY-Zeuthen and the staff of the computer centre at Zeuthen
for their support. 
We are grateful to Hartmut Wittig and Ulli Wolff for discussions
and a critical reading of the manuscript. 
Martin L\"uscher is thanked for numerous discussion and useful
suggestions. 
We are thankful to Roberto
Frezzotti for his help in comparing part of our results with those
obtained with the PHMC algorithm.

\end{document}